\documentclass[aps, superscriptaddress, showpacs, twocolumn, longbibliography, prx, nofootinbib]{revtex4-2}
\usepackage{units}
\usepackage{etoolbox}
\usepackage{amsmath}
\usepackage{amssymb}
\usepackage{graphicx}
\usepackage{hanging}
\usepackage{bm}
\usepackage{multirow,color,relsize,ulem,microtype,mathtools}
\usepackage{appendix}
\usepackage{afterpage}
\usepackage{textgreek}
\usepackage{upgreek}
\usepackage{soul} 
\usepackage{csquotes}
\soulregister\ref{7}
\soulregister\eqref{7}
\soulregister\cite{7}
\soulregister\onlinecite{7}
\usepackage[colorlinks=true, allcolors=blue]{hyperref}
\usepackage[capitalise]{cleveref}

\renewcommand{\st}[1]{}

\newcommand{\be}{\begin{equation}}
\newcommand{\ee}{\end{equation}}


\begin{document}

\title{Modeling Light Propagation and Amplification Efficiency in Highly Multimode, Yb-doped Fiber Amplifiers}

\author{Darcy L. Smith$^{*,\dagger}$}
\affiliation{Institute for Photonics and Advanced Sensing (IPAS) and the School of Physics, Chemistry and Earth Sciences, Adelaide University, 5005, SA, Australia}
\affiliation{Department of Applied Physics, Yale University, New Haven, 06520, CT, USA}
\affiliation{OzGrav - Australian Research Council Centre of Excellence for Gravitational Wave Discovery, Adelaide Node, SA, Australia}

\author{Kabish Wisal$^{*,\dagger}$}
\affiliation{Department of Applied Physics, Yale University, New Haven, 06520, CT, USA}

\author{Baichuan Huang}
\affiliation{Department of Applied Physics, Yale University, New Haven, 06520, CT, USA}

\author{Stephen C. Warren-Smith}
\affiliation{Institute for Photonics and Advanced Sensing (IPAS) and the School of Physics, Chemistry and Earth Sciences, Adelaide University, 5005, SA, Australia}
\affiliation{Future Industries Institute, Adelaide University, Mawson Lakes, 5095, SA, Australia}

\author{Ori Henderson-Sapir}
\affiliation{Institute for Photonics and Advanced Sensing (IPAS) and the School of Physics, Chemistry and Earth Sciences, Adelaide University, 5005, SA, Australia}
\affiliation{OzGrav - Australian Research Council Centre of Excellence for Gravitational Wave Discovery, Adelaide Node, SA, Australia}

\author{Hui Cao}
\affiliation{Department of Applied Physics, Yale University, New Haven, 06520, CT, USA}

\author{David J. Ottaway}
\affiliation{Institute for Photonics and Advanced Sensing (IPAS) and the School of Physics, Chemistry and Earth Sciences, Adelaide University, 5005, SA, Australia}
\affiliation{OzGrav - Australian Research Council Centre of Excellence for Gravitational Wave Discovery, Adelaide Node, SA, Australia}

\author{A. Douglas Stone}
\affiliation{Department of Applied Physics, Yale University, New Haven, 06520, CT, USA}

\begin{abstract}

Multimode fibers have been proposed for mitigating nonlinear effects in high-power fiber amplifiers, allowing for significant power scaling. Most previous studies on light propagation in continuous-wave fiber amplifiers focus on single mode or few mode fibers. Here we develop a tractable numerical model to simulate light propagation in narrowband, highly multimode fiber amplifiers, which takes into account gain saturation, pump depletion and mode-dependent gain. We consider a frequency domain, field based model, with modal gain being dependent on both intramodal gain and gain-induced mode coupling. We derive coupled equations for the evolution of signal modal amplitudes, pump power and population inversion, and numerically solve these equations using a finite-difference method. For highly multimode excitations, the optical intensity in the fiber is speckled and various modes grow at different rates, due to differential overlap with the gain medium and spatial hole burning. Our analysis is applied to Yb-doped fibers, with a quasi-quantitative analysis of the specific case of Yb, identifying different regimes in which either spontaneous emission (SE) or amplified spontaneous emission (ASE) limit amplifier efficiency, especially for larger core and multimode fibers. Finally, we incorporate ASE and spectrally resolved optical channels into our model and demonstrate the experimentally verifiable phenomenon of ASE suppression with sufficient input signal power. Our model can be combined with existing models for various nonlinear effects, providing a useful tool for quantitatively studying nonlinearity mitigation and power scaling in multimode fiber amplifiers.

\end{abstract}

\maketitle

\begingroup
\renewcommand\thefootnote{}
\footnotetext{
$^{*}$ Corresponding authors: darcy.smith@adelaide.edu.au, kabish.wisal@yale.edu \\
$^{\dagger}$ These authors contributed equally to this work.
}
\addtocounter{footnote}{0}
\endgroup

\section{Introduction}

High-power fiber lasers utilizing multi-stage amplifiers enable a compact platform to achieve ultra-high laser power \cite{jauregui2013HPFLs,zervas2014HPFLs,dawson2008HPFLs,hecht18HPFLs}. This has created a potential for applications in both fundamental science such as in gravitational wave detection \cite{wellmann19gravwave}, and industrial applications such as in advanced manufacturing \cite{carter19welding}, LiDAR \cite{liu2016lidar}, and defense \cite{kaushal2017defense}. To fully realize this potential, further power scaling is required. This is primarily limited by various nonlinear effects such as stimulated Brillouin scattering (SBS) \cite{agrawal2019nonlinear,wolff2021SBS,kobyakov2010SBS}, transverse mode instability (TMI) \cite{jauregui2020tmi,smith2011tmi,zervas2018tmi}, and possibly in the future Kerr-induced modulation instability \cite{dupiol2017MI,agrawal1989MI}. The mitigation of SBS and TMI in fibers is an ongoing area of research closely related to the challenge of power scaling in fiber lasers. Specifically for amplifiers with narrow linewidth, a requirement for important applications such as coherent beam combination and LiDAR, mitigating both SBS and TMI has been difficult. Most previous approaches utilize single mode fibers (SMF), motivated primarily by the high beam quality afforded by the Gaussian-like fundamental fiber mode. As such, most of the modeling efforts in fiber amplifiers have focused on SMF \cite{giles1991EDFAs,paschotta1997ytterbium,agrawal1990amplification}. 

In several recent theoretical and experimental studies it was shown that both SBS and TMI can be efficiently mitigated by coherent multimode excitation in highly multimode fibers (MMFs)~\cite{ke2014sbs,chen2023wavefrontshaping,wisal2024nonlinear,wisal2024TMI,wisal2024sbs,chen2023TMI}. It has also been shown that for coherent excitations, a diffraction limited focused output beam can be obtained by shaping the wavefront at the input in both passive and active fibers \cite{florentin2017activeMMF,rothe2024beamshaping,gomes2022focussing,chen2025,rothe2025wavefrontshaping}. This demonstrates that MMFs can provide a novel platform for mitigating nonlinearities and achieving significant power scaling in fiber amplifiers. To study these detrimental nonlinear effects in MMF amplifiers quantitatively, an accurate numerical model of light propagation and gain-induced nonlinearities in fibers under arbitrary multimode excitations is necessary. Another reason for significant recent interest in MMFs has been to study the interplay between nonlinear effects and spatial degrees of freedom, such as in Kerr beam self-cleaning \cite{krupa2017selfcleaning}. In this work we provide a numerical model of amplification of narrowband light in MMFs, which takes into account gain saturation, pump depletion and mode-dependent gain. Our model of MMF amplifiers can then be combined with models for various unwanted nonlinear effects such as SBS, TMI and Kerr-induced four wave mixing, allowing quantitative predictions.

As mentioned above, significant work has been done in modeling light amplification in SMF amplifiers. The physical origin of the amplification of light is stimulated emission as it propagates through an optically pumped gain medium, which typically consists of rare-earth elements such as, Yb, Er, etc. doped into a silica fiber. Numerical models of fiber amplifiers require solving the optical wave equation for the signal coupled to pump light through the gain medium. The local gain is computed from the spatially-varying population inversion by solving steady-state rate equations. A key assumption commonly used is that the transverse mode profile of the amplified signal within the fiber is constant. This reduces the problem to a set of one-dimensional differential equations describing the evolution of pump and signal power, along with the population inversion. This framework accurately models light propagation in SMF, however, it becomes inaccurate if light propagates in multiple fiber modes. In particular, it fails to capture mode coupling due to spatial hole burning, which generates rapid spatial variations in the local gain due to multimode interference and gain saturation. 

To accurately model a MMF amplifier, light propagation needs to be considered in multiple transverse modes, modeled using field amplitudes to account for multimode interference. In addition, mode-dependent gain and loss, and mode coupling induced by the presence of the gain medium within the fiber core, must be considered. An intensity-based model for MMF amplifiers was developed in Ref.~\cite{kan2012intensity}, and while it was suitable for the study of few-moded fiber amplifiers (as used in telecommunications applications), could not be accurately extended to more multimoded fibers, as it neglected modal interference, and hence leads to issues with accurately capturing power conservation and calculating efficiency. An improved model, involving multimode field amplitudes, was presented in Ref.~\cite{trinel2017theoretical}, however the scope of this work was also few-moded telecommuncations fibers, not highly MMFs. Another interesting study, which modeled MMF amplifiers, involved numerical simulation of the nonlinear Schr{\"o}dinger equation to describe non-linear effects in MMF amplifiers~\cite{chen2023}. Since, both Refs.~\cite{trinel2017theoretical} and \cite{chen2023} involved broadband light consisting of multiple frequencies, they were computationally limited to few-moded fibers. Mathematically, the computational runtime for such models scales as $\mathcal{O}(N_wN_o^2)$, where $N_w$ is the number of frequencies and $N_o$ is the number of modes \cite{wright2018computational}. 

For studying high-power, \textit{narrow-linewidth}, MMF amplifiers, it is possible to develop an efficient computational model, even for \textit{highly multimode} excitations. In this paper, we present an efficient yet accurate numerical model for MMF amplifiers, which involves a multimode single-frequency signal, along with broadband and incoherent pump and ASE light. If the signal linewidth is significantly smaller than the gain bandwidth, the signal can be considered as single-frequency. Furthermore, for an incoherent pump source and broadband ASE, speckle behavior can be neglected and simplified to a uniform transverse profile. Hence the computational runtime in our approach scales as $\mathcal{O}(N_o^2+N_w)$, simultaneously allowing for a highly multimode signal and a well resolved, broadband ASE spectrum without unreasonable computational costs. Our assumptions about the signal, pump and ASE light are informed by the experimental configurations commonly used in high-power MMF amplifiers with quasi-monochromatic continuous-wave (CW) output. 

We begin by presenting a comprehensive derivation of coupled equations for the evolution of signal modal amplitudes, pump power and fractional upper-level population. We solve these equations numerically with a fourth order Runge-Kutta finite difference method. Initially, we ignore ASE, and only consider multimode signal growth along with the evolution of pump. Later, we consider a broadband model incorporating ASE. We simulate a highly multimode fiber at $1064 \: \rm nm$ signal wavelength with a Yb-doped gain medium. We present results for the total signal power, pump power, individual modal power, modal gain, spatial-hole burning, signal reabsorption and pump-signal conversion efficiency for the first time with a fully field-based, multimode model. Crucially, we incorporate the cross-gain terms resulting from spatially varying gain saturation and population inversion. These terms are required to accurately model energy conservation and efficiency calculations. This is the only source of mode coupling; non-universal linear and nonlinear mode coupling due to the variations in the real part of the refractive index are neglected as they leave overall behavior qualitatively unchanged. Finally, we generalize our model to include broadband ASE and study output efficiency, evolution of the gain spectrum and output beam profile as a function of signal power. Overall, our work contributes an important tool for studying multimode fiber amplifiers, providing a promising new platform for instability-free ultra high laser power.

\section{Multimode fiber amplifier model}

\subsection{Multimode signal growth equations}
\label{ModalGainDerivation}

We begin by deriving growth equations for signal light propagating through an active MMF, which are then solved numerically. We decompose the signal light into a set of transverse eigenmodes, each a solution to the optical wave equation given the fiber geometry. Previously, in the development of intensity-based MMF amplifier models, it was assumed that each mode propagates and interacts with the gain medium as an independent channel \cite{kan2012intensity,giles1991EDFAs}. However, this assumption is not accurate, and the error associated with such a model grows with the number of propagating modes. In our work, we do not make this assumption, and show that the amplitude equations for various modes are coupled by the presence of the spatially-dependent gain medium.

In this section, the coupled equations governing the evolution of each modal amplitude and the gain-induced inter-modal coupling is derived from first principles. It is possible to arrive at these equations by combining multiple results in published work such as in Refs.~\cite{trinel2017theoretical,chen2023}. However, a clear and comprehensive prior derivation is lacking, to our knowledge. As such, in this section our intent is both pedagogical, to fill this gap, and to reach a starting point for the numerical results provided in this paper. 

We consider a rare-earth doped fiber amplifier pumped from the input end (co-pumping). Later we also treat pumping from the output end (counterpumping). A broadband, incoherent pump beam excites the rare-earth ions, creating a population inversion. A coherent multimode signal beam is injected into the fiber core and undergoes amplification through stimulated emission. The electric field for the signal $\textbf{E}_s$ obeys the Helmholtz electromagnetic wave equation:
\begin{gather}
\left[\nabla^2-\frac{n^2}{c^2}\frac{\partial^2}{\partial t^2}\right]\mathbf{E}_s=0 , \label{WaveEquation}
\end{gather}
where, $c$ is the speed of light in vacuum. $n$ is the local refractive index, which takes the following form:
\begin{gather}
n(\mathbf{r})=n_0(\mathbf{r}_\perp)+ in_g(\mathbf r),
\end{gather}
where $n_0(\mathbf{r}_\perp)$ is the refractive index profile of the fiber in the absence of active dopants, assumed translationally invariant, and $in_g(\mathbf r)$ is the change in the index of refraction due to the presence of the gain medium. A cylindrical coordinate system is used with $z$ being aligned with the fiber axis; $\mathbf{r}_\perp$ represents the combined coordinates $(x,y)$ in the plane transverse to the fiber axis. The imaginary part represents the processes of absorption/emission provided by the gain medium ($n_g$ is negative in regions and at frequencies where the pumping creates gain, and positive elsewhere). The frequency-dependent change in the real part of $n$ due to $n_g$ as dictated by the Kramers-Kronig relations is neglected. This is because this term only redistributes power within the fiber modes, leaving amplifier behavior qualitatively unchanged. In addition, the specific value of change in the real part of the refractive index due to the gain medium is highly specific to a given fiber along with the environmental conditions. For similar reasons, we also neglect random variations in the refractive index profile due to fiber imperfections leading to linear mode coupling. Both of these effects can be incorporated in our formalism in a  straightforward manner if needed, based on experimental characterization of the particular fiber of interest.

We assume that $n_g\ll n_0$, such that $n^2$ can be approximated as:
\begin{gather}
n^2\approx n_0^2+2in_0n_g .
\end{gather}
Additionally, we expand the electric field in terms of the guided modes of the fiber:

\begin{gather}
\mathbf{E}_s(\mathbf {r})=\sum_mA_m(z) \boldsymbol{\psi}_m(\mathbf{r}_\perp)e^{i(\beta_mz-\omega_0t)} ,
\end{gather}
where, $\omega_0$ denotes the signal frequency, $A_m(z)$ denotes the modal amplitude, $\boldsymbol{\psi}_m(\mathbf{r}_\perp)$ denotes the transverse mode electric-field profile, and $\beta_m$ denotes the propagation constant for the $m^{\rm th}$ mode. Each fiber mode satisfies the fiber modal equation given by~\cite{snyder1983optical},

\begin{gather}
\left[\nabla_T^2+\left(\frac{n_0^2\omega_0^2}{c^2}-\beta_m^2\right) \right]\boldsymbol{\psi}_m=0. \label{ModalEigenvalue}
\end{gather}
Here, $\nabla_T^2$ is the transverse Laplacian. Substituting our expressions for $n$ and $\mathbf E$ into \cref{WaveEquation}, we obtain:

\begin{gather}
\left[\nabla^2+\frac{n_0^2\omega_0^2}{c^2}+i\frac{2n_0n_g\omega_0^2}{c^2}\right]\nonumber\\\times\sum_mA_m(z)\boldsymbol{\psi}_m(\mathbf{r}_\perp)e^{i\beta_mz}=0 .
\end{gather}
We split the Laplacian operator into its transverse and longitudinal components, apply the slowly varying envelope approximation, and utilize the modal eigenvalue equation given in \cref{ModalEigenvalue}, obtaining:

\begin{gather}
\mathlarger{\mathlarger{\sum}}_m2i\beta_m\frac{d A_m(z)}{d z}\boldsymbol{\psi}_m(\mathbf{r}_\perp)e^{i\beta_mz}\nonumber\\=\frac{2in_0n_g\omega^2}{c^2}\sum_nA_n(z)\boldsymbol{\psi}_n(\mathbf{r}_\perp)e^{i\beta_nz} .
\end{gather}
To isolate a single term corresponding to mode $m$ on the left hand side (LHS), we take the inner product with $\boldsymbol{\psi}_m^*$ on both sides and integrate across the transverse cross-section of the fiber, leveraging the orthonormality of transverse fiber eigenmodes. With some rearrangement, this leads to:
\begin{gather}
\frac{d A_m(z)}{d z}=\frac{n_0\omega_0^2}{\beta_mc^2}\sum_nA_n(z)e^{i(\beta_n-\beta_m)z}\nonumber\\\times\int n_g(\mathbf{r}_\perp,z)\:\boldsymbol{\psi}_m^*.\boldsymbol{\psi}_n\:d\mathbf{r}_\perp \:. \label{ModalGain1}
\end{gather}

\noindent
It can be seen that the change in amplitude of mode $m$ with respect to propagation along $z$ depends on the amplitudes of all fiber modes. The reason for the coupling between various modal amplitudes is the presence of transverse spatial dependence in the refractive index $n_g(\mathbf{r}_\perp,z)$. If $n_g$ were constant across the fiber cross-section, the integral on the right hand side (RHS) of \cref{ModalGain1} would become proportional to $\delta_{mn}$, decoupling the growth for each modal amplitude. This is the assumption used in previous intensity-based models of MMF amplifiers, leading to inaccuracies. However in the presence of gain saturation, $n_g$ depends on the local signal intensity, which in MMFs is highly speckled, leading to strong spatial variations in $n_g$. As such, a coupled treatment of the modal amplitudes is necessary for accurate and consistent results. 

Next, we write $n_g$ in terms of the signal gain coefficient $g(\mathbf{r})$ due to stimulated emission, which depends on the population inversion~\cite{siegman1986lasers,hecht2017optics}:
\begin{gather}
2k_0n_g(\mathbf{r})= g(\mathbf{r}) = \sigma_{e,s}N_2(\mathbf{r})-\sigma_{a,s} N_1(\mathbf{r}),
\label{gcoeff}
\end{gather}
where $k_0=\frac{\omega_0}{c}$, $\sigma_{e,s}$ and $\sigma_{a,s}$ represent the emission and absorption cross-section at the signal frequency, and $N_2$ and $N_1$ are the number volume density of active ions in the excited and ground state respectively. The above expression for $n_g$ allows us to write the equation for modal amplitude evolution in the following form:
\begin{gather}
\frac{d A_m(z)}{d z}=\sum_n g_{mn}(z) A_n(z) e^{i(\beta_n-\beta_m)z}, \label{ModalGain2}
\end{gather}
where,
\begin{gather}
g_{mn}(z)=\frac{1}{2}\int \boldsymbol{\psi}_m^*.\boldsymbol{\psi}_n\: g(\mathbf{r}_\perp,z) \:d\mathbf{r}_\perp . \label{gmn}
\end{gather}

\noindent
The above two equations describe the multimodal signal growth and their solution forms a major part of our model. In the above form, these equations have a rather straightforward interpretation. The amplitude growth in any mode depends linearly on the amplitudes in each mode, with a corresponding gain coefficient between any two modes $g_{mn}$ depending on the overlap between the modal profiles and the local signal gain coefficient $g(\mathbf{r})$, which depends on the local population inversion as given by \cref{gcoeff}. In the case of linear gain, where $g$ is approximately constant in space, the equation for each modal amplitude becomes independent, leading to exponential growth in power in each mode, as expected. However, in general, in high-power fiber amplifiers, the local population inversion depends substantially on the local signal and pump intensity, making the signal growth equations highly nonlinear. As a result, finding analytical solutions to the signal growth equations is not possible, except in some very special cases. As such, in our model, we solve the signal growth equations numerically using a finite difference method. Before proceeding to that solution, we need to describe how the local population inversion, characterized by $N_1$ and $N_2$, depends on the signal and the pump intensity. 

\subsection{Population inversion and pump depletion}

In our model, we consider a MMF amplifier that is doped with a rare-earth metal such as Ytterbium (Yb) or Erbium (Er), which provides signal gain via stimulated emission. In general, these ions have three energy levels which participate in amplification, as well as additional energy levels which may participate in processes such as excited-state absorption or upconversion. It has been shown however that these three-level systems can be modeled as effectively two-level systems owing to a separation in time scales associated with various transition rates~\cite{giles1991EDFAs,desurvire2002erbium}. In all cases in this paper, the gain medium is treated as such a two-level medium. In the case that a gain medium has a level structure which requires a more complicated treatment, the modeling of population inversion can be generalized without needing to change the signal and pump evolution equations in our model.

For simplicity, we assume that the dopant ions are uniformly spread over the fiber core with density $N_{tot}$. These ions occupy either the lower or the upper energy level, with population densities $N_1$ and $N_2$ respectively, such that $N_1+N_2=N_{tot}$. Stimulated transitions between these levels are facilitated by the presence of pump light at frequency $\nu_P$ and signal light at frequency $\nu_S$. Following the derivation presented in Refs.~\cite{giles1991EDFAs,desurvire2002erbium,paschotta1997ytterbium}, the density of ions in the upper level in the steady-state regime is given by:
\begin{gather}
N_2 = N_{tot}\frac{\frac{\sigma_a(\nu_p)}{\sigma_a(\nu_p)+\sigma_e(\nu_p)}\frac{I_p(\mathbf{r}_\perp,z)}{I_{\rm sat}(\nu_p)}+\frac{\sigma_a(\nu_s)}{\sigma_a(\nu_s)+\sigma_e(\nu_s)}\frac{I_s(\mathbf{r}_\perp,z)}{I_{\rm sat}(\nu_s)}}{1+\frac{I_p(\mathbf{r}_\perp,z)}{I_{\rm sat}(\nu_p)}+\frac{I_s(\mathbf{r}_\perp,z)}{I_{\rm sat}(\nu_s)}}, \label{N2}
\end{gather}
where saturation intensity is given by
\begin{gather}
I_{\rm sat}(\nu)=\frac{h\nu}{(\sigma_a(\nu)+\sigma_e(\nu))\tau}\label{Isat}.
\end{gather}

\noindent
Here, $\sigma_a(\nu)$ and $\sigma_e(\nu)$ represent the absorption and emission cross-section at frequency $\nu$, with $\nu=\nu_s$ denoting the signal frequency and $\nu=\nu_p$ denoting the pump frequency. Upper level population density $N_2$ depends on both the signal intensity $I_s$ and the pump intensity $I_p$ relative to the respective saturation intensities $I_{\rm sat}(\nu_s)$ and $I_{\rm sat}(\nu_p)$. Note that the saturation intensity depends on the energy of the pump or signal photon divided by the total absorption and emission cross-section and the upper level lifetime $\tau$, the inverse of the spontaneous emission (SE) rate. As written in \cref{N2} various terms in the denominator represent rates of both emission and absorption of the signal and the pump  normalized to the rate of SE. Similarly, the numerator consists of terms representing the relative rate of only the absorption of the pump and the signal. Overall, the ratio of the numerator (representing absorption) and the denominator (representing both absorption and emission) gives the steady state population of the upper level. A derivation of this steady-state population inversion equation is given in \cite{desurvire2002erbium}.

Given $N_{tot}$ as a known parameter of the fiber, \cref{N2} can be used to calculate $N_2$ and $N_1$ at any point in the fiber core, if the signal intensity $I_s$ and pump intensity $I_p$ are known. The remaining quantities depend on material and optical constants. The signal intensity $I_s(\mathbf{r}_\perp,z)$ is calculated from a coherent sum of the fiber modes, given by
\begin{gather}
I_s(\mathbf{r}_\perp,z) = \left|\mathbf\sum_mA_m(z)\boldsymbol{\psi}_m(\mathbf{r}_\perp)e^{i\beta_mz} \right|^2.
\end{gather}

Most standard fiber amplifiers are cladding-pumped, in which a broadband pump is launched into the pump core (the fiber inner cladding + core), while the dopant ions which absorb the pump are present only in the core. As the pump source is usually a spectrally incoherent diode launched into the many cladding modes, its speckle pattern `smears out', and the its transverse spatial profile over the fiber core can be assumed as constant. The pump intensity is then given as $I_p(z) = P_p(z)/A_{cl}$, where $P_p(z)$ is the total pump power at location $z$ along the fiber axis and $A_{cl}$ is the area of the inner cladding. Although the pump intensity is constant across the fiber cross-section, it varies significantly in the longitudinal direction, as the pump power decreases along the fiber axis due to pump absorption, usually referred to as pump depletion. Based on the rate equations, the evolution of pump power is given by \cite{desurvire2002erbium}:

\begin{gather}
\nonumber\frac{dP_P(z)}{dz}=\\-\frac{P_P(z)}{A_{cl}}\int d\mathbf{r}_\perp(\sigma_{a,p}N_1(\mathbf{r}_\perp,z)
-\sigma_{e,p}N_2(\mathbf{r}_\perp,z)). \label{PumpAbsorption}
\end{gather}
With a negligible emission cross-section at the pump frequency, the pump decreases exponentially, with the local pump absorption coefficient depending on the population inversion.

\subsection{Numerical solution approach}

To summarize, our model consists of amplitude growth equations for the individual modal amplitudes of the signal (\cref{ModalGain2} and \cref{gmn}), and a single equation describing pump power depletion (\cref{PumpAbsorption}), coupled by a nonlinear steady state formula for the local population inversion (\cref{N2}). Together these equations fully describe the coherent amplification of a narrowband multimode signal in the presence of dopant ions excited by broadband, transverse spatially uniform pump light. One major effect we have not considered in this section is ASE, which can also extract energy from the pump. This will be discussed in \cref{BroadbandSection}.

Since the equations for the signal, population inversion and the pump are coupled and nonlinear, solving them analytically is not possible. To overcome this, we develop a numerical scheme for solving these equations based on the finite-difference method. Our algorithm can be summarized as follows. At the fiber input, we are given a pump power $P_p(z=0)$ and a multimode signal, represented by a set of complex modal amplitudes $\{A_m(z=0)\}$. The transverse mode profiles $\{\boldsymbol{\psi}_m \}$ and propagation constants $\{\beta_m\}$ are calculated and stored for the given fiber. Assuming all the relevant material and optical constants are known, the goal is to calculate the signal, pump and population inversion throughout the fiber, i.e., obtain $\{A_m(z)\}$, $P_p(z)$, and $N_2(\mathbf{r}_\perp,z)$. This is achieved as follows.

Say at a point $z$, $\{A_m(z)\}$ and $P_p(z)$ are known, then we calculate $N_2(\mathbf{r}_\perp,z)$ using \cref{N2}. With $N_2$ (and $N_1=N_{\rm tot} - N_2$), we calculate the local signal gain $g_{mn}$, providing the value of the RHS in \cref{ModalGain2} and \cref{PumpAbsorption}. By discretizing the derivative on the LHS in \cref{ModalGain2} and \cref{PumpAbsorption}, we update the value of $A_m(z+\Delta z)$ and $P_p(z+\Delta z)$ using the value of variables and derivatives at the previous steps. In this manner, we continue to iterate forward along the length of the fiber until $z=L$, alternating between calculating $N_2$ using the signal and pump intensity and updating the signal amplitudes and the pump power at the next step using the signal gain and pump absorption obtained from $N_1$ and $N_2$. The precise rule for updating depends on the discretization scheme employed. In our case, we considered both the Euler method and a fourth-order Runge-Kutta method. For sufficiently small $\Delta z$, both methods were able to produce accurate and consistent results. In general, Runge-Kutta provides higher accuracy but the Euler method requires less computational runtime, in accordance with the previous literature. The results presented in this work used Runge-Kutta.

$\Delta z$ is determined by considering the smallest longitudinal speckle-based feature size, given by $L_{speckle}=\frac{2\pi}{\beta_{max}-\beta_{min}}$, and suitably resolving this feature with at least 20 longitudinal points.

\subsection{Results of the single-frequency MMF amplifier model} \label{SingleFrequencySection}

\begin{figure*}[!t]
\includegraphics[width=\textwidth]{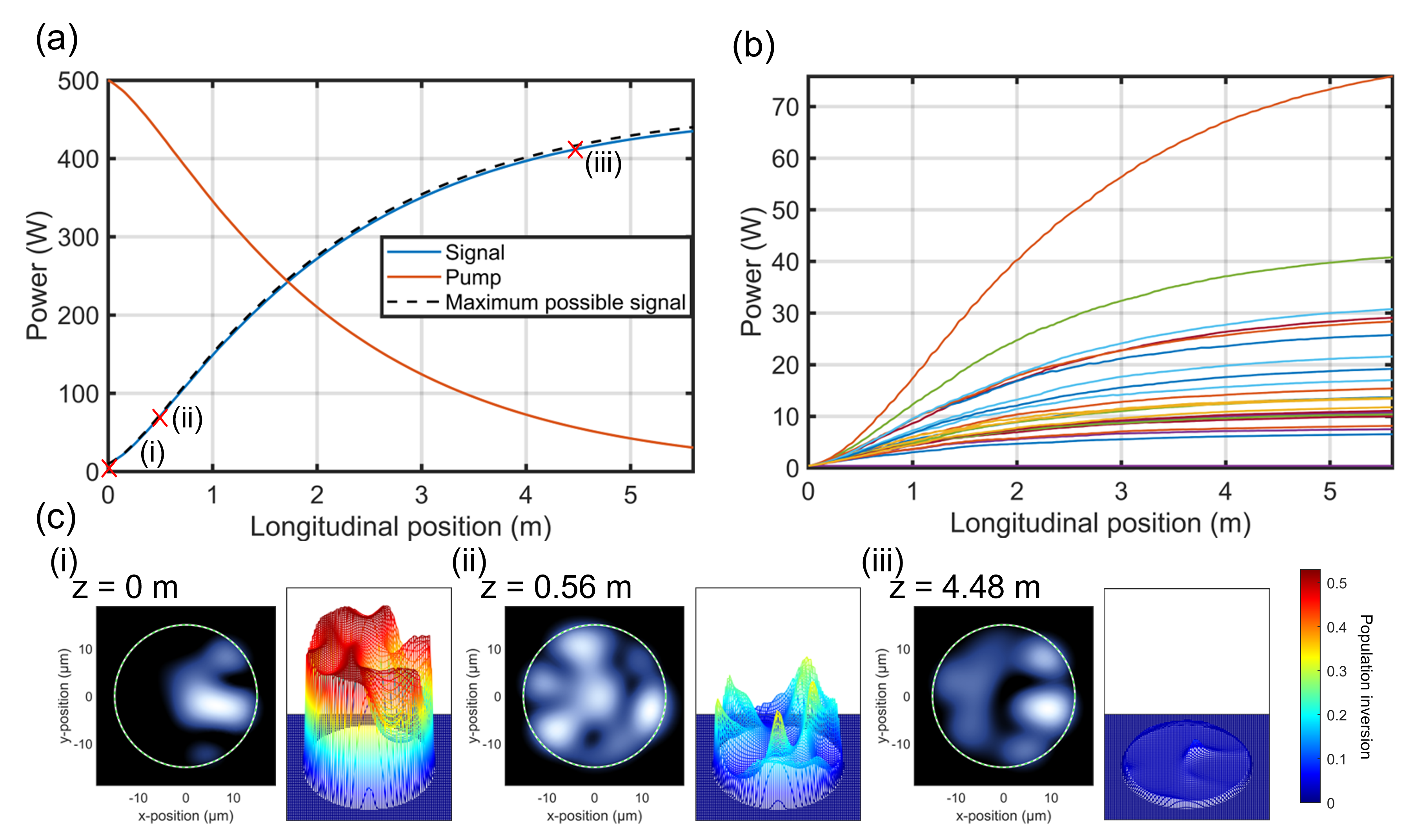}
\centering
\caption{Results of the multimode fiber amplifier model. 500 W of pump power and 10 W of signal power, distributed evenly amongst the 24 supported modes (accounting for the two polarizations corresponding to each propagation constant), are launched into the fiber described in \cref{ParametersA}. (a) Signal and pump power as a function of z. The maximum possible signal, as set by the Stokes efficiency, is displayed with the dashed line. (b) Power carried by each individual mode as a function of z. (c) Transverse multimode speckle pattern (2D plot on left) and resulting population inversion (3D mesh plot on right) at (i) z = 0 m, (ii) 0.56 m and (iii) 4.48 m (indicated on (a)).}
\label{SingleFrequencyResults}
\end{figure*}

\begin{figure*}[!t]
\includegraphics[width = \textwidth]{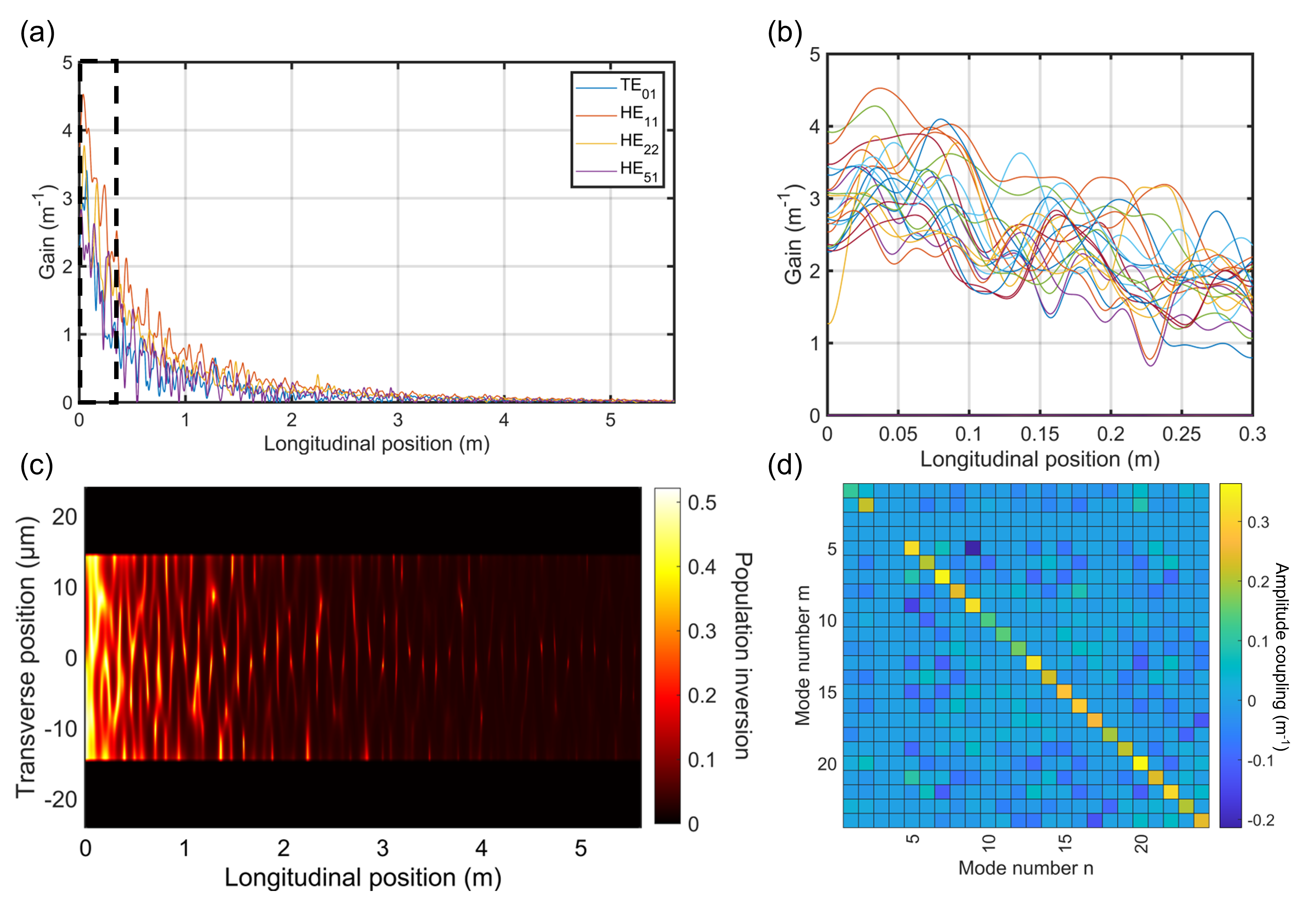}
\centering
\caption{(a) Modal gain as a function of z for four selected modes from the fiber described in \cref{ParametersA}. (b) Modal gain for all 24 supported modes of the fiber, zoomed-in to the inset section of (a). (c) Slice of the population inversion along the x-z plane. (d) The $g_{mn}$ matrix, as defined in \cref{gmn}, at z=5.6 m. }
\label{GainResults}
\end{figure*}

\begin{table}[h!]
    \centering
    \begin{tabular}{@{} l c @{}}
        \hline\hline
        \textbf{Parameter} & \textbf{Value} \\
        \hline
        Core diameter ($\upmu$m) & 30 \\
        Cladding diameter ($\upmu$m) & 420 \\
        Core refractive index & 1.447 \\
        Cladding refractive index & 1.445 \\
        Core Yb population density (m$^{-3}$) & $10^{26}$ \\
        Fiber length (m) & 5.6 \\
        Signal wavelength (nm) & 1064 \\
        Pump wavelength (nm) & 975 \\
        Number of supported modes & 24 \\
        \hline\hline
    \end{tabular}
    \caption{Parameters of the fiber used to produce the results in \cref{SingleFrequencyResults,GainResults}} \label{ParametersA}
\end{table}
The parameters used for the Yb-doped MMF amplifier are given in \cref{ParametersA}, representing realistic parameters for a 24 mode amplifier (including both polarisations), and the results are shown in \cref{SingleFrequencyResults}. The steady-state pump and signal power as a function of $z$ is presented in \cref{SingleFrequencyResults}(a). The pump power decreases as the pump is absorbed and the signal power grows by extracting energy from the gain medium, as expected. The signal power curve is comprised of a sum of each individual mode's power evolution, shown in \cref{SingleFrequencyResults}(b). The multimode speckle of the signal and the resulting transverse population inversion distribution are shown at three points along the fiber in \cref{SingleFrequencyResults}(c), demonstrating the progression from a highly excited to a depleted gain medium, as well as the transverse spatial hole burning induced by the multimode speckle.

While the total signal power evolution shown in \cref{SingleFrequencyResults}(a) and even the modal power evolution in \cref{SingleFrequencyResults}(b) are relatively smooth, the gain experienced by each mode exhibits a strongly noisy oscillatory pattern as exemplified by \cref{GainResults}(a) and (b). This is due to a combination of two multimode effects - the effect of speckle on the population inversion at a given $z$ (as predicted in \cref{N2} and exhibited in \cref{GainResults}(c)), and the individual overlap of each mode with this resulting population inversion which induces coupling between modes (as predicted in \cref{gmn} and exhibited in \cref{GainResults}(d)). The result is a random walk-like behavior in modal gain, with an overall envelope determined by the pump and signal powers, and deviations from this determined by the speckled nature of the multimode propagation.

\begin{table}[h!]
    \centering
\begin{tabular}{lc}
    \hline\hline
    \multicolumn{2}{c}{\textbf{Yb parameters}} \\
    \hline
    Parameter & Value \\
    \hline
    $\sigma_{a,s}$ (m$^2$) & $5.8 \times 10^{-27}$ \\
    $\sigma_{e,s}$ (m$^2$) & $2.71 \times 10^{-25}$ \\
    $\sigma_{a,p}$ (m$^2$) & $1.19 \times 10^{-24}$ \\
    $\sigma_{e,p}$ (m$^2$) & $1.02 \times 10^{-24}$ \\
    $\tau$ (s) & $9.01 \times 10^{-4}$ \\
    $\lambda_s$ (nm) & 1064 \\
    $\lambda_p$ (nm) & 975 \\
    \hline\hline
\end{tabular}
    \caption{Relevant parameters of Ytterbium.}
    \label{YbErTable}
\end{table}
 


It can be seen from \cref{SingleFrequencyResults} that at all points in the fiber, the total signal power is less than the `maximum possible signal', defined as

\begin{gather}
P_{S,max}(z) = P_S(0)+\eta_{max}(P_P(0)-P_P(z)) ,
\end{gather}

where the Stokes efficiency ($\eta_{max}$) is defined as $\lambda_P/\lambda_S$. $P_{S,max}(z)$ gives the signal power at a point $z$ in the fiber if every pump photon absorbed up to that point resulted in a coherently produced signal photon through stimulated emission. As the signal transition is of a lower energy than the pump transition, this is less than 100\% of the absorbed pump, with the remaining energy going into the quantum defect heating. Thus the maximum possible signal power is set by the Stokes efficiency, which for the pump and signal wavelengths used here is $\sim 92\% $. All quoted efficiencies in this paper will be relative to $\eta_{max}$, and hence it is important to note that efficiencies must be multiplied by $\eta_{max}$ to convert them to pure optical efficiency (such as slope efficiencies quoted in experimental works).

The Stokes efficiency only represents the upper limit on amplifier efficiency. Beyond the power lost through non-radiative energy dissipation, power is also lost through isotropically emitted spontaneous emission (SE), due to the finite metastable state lifetime $\tau$.  In addition, the small fraction of the SE emitted into guided modes is amplified, leading to broadband ASE, which competes with the signal for pump power and reduces the amplifier efficiency. The modeling of ASE is more complicated as it occurs everywhere within the fiber and can be guided in the backwards as well as forward direction. The explicit effect of ASE on the amplifier behavior will be discussed in detail in \cref{broadband}. However, qualitative analyses on regimes of amplifier behaviour as it is limited by SE and ASE can be made by studying the expression for population inversion given in \cref{N2}.

\begin{figure*}[!t]
\includegraphics[width = \textwidth]{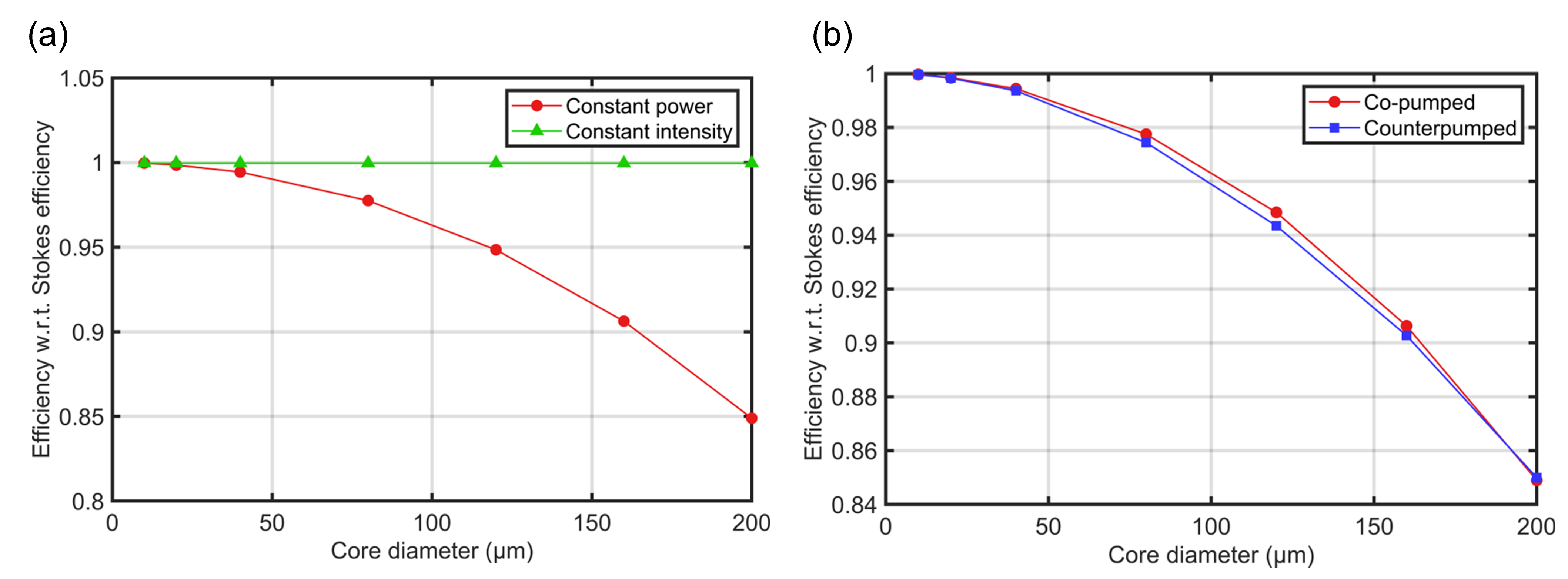}
\centering
\caption{(a) Amplifier efficiency with respect to the Stokes efficiency as a function of core diameter (with the cladding diameter scaled by a constant factor of the core diameter - here $d_{cl}=5d_{co}$). Nominal power is defined as 1 kW input pump and 30 W input signal at 10 $\upmu$m core diameter, with constant intensity demanding this power is increased as per the square of the core size increase. Dopant density is scaled as 1/$d_{co}^2$ such that the dopant number remains constant as core size is changed. In this way, the pump absorption length remains similar for all cases. The first 20 modes of each fiber are chosen for fibers supporting more than 20 modes for comparison and for computational reasons. (b) Amplifier efficiency with respect to the Stokes efficiency for the cases of co- and counterpumping, with parameters otherwise the same as the red curve in (a).}
\label{Efficiency}
\end{figure*}

We can rewrite \cref{N2} in the form:
\begin{gather}
N_2 = N_{tot}\frac{f_a^p S_p/\tau + f_a^s S_s/\tau} {(\frac{1}{\tau} + \frac{S_p}{\tau}  + \frac{S_s}{\tau}) }, \label{N?}
\end{gather}
where $f_a^p,f_a^s$ are the absorption cross-sections for the pump and signal, normalized by the total cross-section and $S_p,S_s$ are the pump and signal intensities normalized to their saturation values:

\begin{gather}
f_a^i=\frac{\sigma_a(\nu_i)}{\sigma_a(\nu_i)+\sigma_e(\nu_i)}\nonumber \\
S_i(\mathbf{r}) = \frac{I_i(\mathbf{r})}{I_{sat}(\nu_i)}\nonumber \\
where\: i=S,P . \label{definitions}
\end{gather}

Note that we have multiplied by the spontaneous emission rate, $1/\tau$, in numerator and denominator. One sees that in principle both strong pumping and strong signal contribute to the population inversion. However, for Yb, $f_a^p \approx 1/2 $, while $f_a^s \sim 10^{-2}$, and $I_{sat}$ is roughly ten times lower for the pump than for the signal. These differences imply that the second term in the numerator is much smaller for Yb and, in most cases, is negligible within the depletion length of the pump. Therefore, the gain is mainly determined by the pump saturation. $S_p,S_s,N_2$ vary in space and need to be calculated in order to model the amplifier, but we will neglect this effect for the current qualitative arguments.

First, as the emission cross-section at the signal frequency is much greater than the absorption cross-section, the system can have gain when $N_2 \approx (\sigma_{a,s}/\sigma_{e,s}) N_1\ll N_1$ (see the expression for gain given in \cref{gcoeff}). In this limit $N_{tot} \approx N_1$ and since $f_a^p$ is order unity, the pump does not need to be saturated for the amplifier to have some relatively weak gain. When this is the case, the first term in the denominator, representing SE, will dominate over the second and third terms, and most of the energy from the pump will be dissipated through SE. As the SE is emitted isotropically, very little of this emission is coupled into the fiber, and little pump power is converted into either signal or ASE guided within the fiber. Therefore it is necessary to saturate the pump at the input to have a useful amplifier. This is indeed the case in most experimental configurations involving high-power fiber amplifiers.

When the pump is saturated ($S_p \gg 1$), but the seed is too weak, then the signal will not be saturated near the input ($S_s \ll 1)$. In this case the second (pump absorption) term in the denominator dominates, yielding $N_2 \sim N_1$, and the gain will approach its maximum possible value. However, SE will still represent an important loss mechanism, as it occurs more often than stimulated emission of the signal (third term in the denominator) near the input. Moreover, as already noted, the small fraction of SE which couples into the guided modes of the amplifier can be amplified by the pump as  ASE.  However, ASE will only be dominant over SE near the input when this fraction, $ 2\Delta  \Omega/4\pi \sim 10^{-2}> S_s$ (where $\Delta \Omega$ is the solid angle subtended by the guided modes - see Eq. S.22). For Yb, we find that this only occurs for a relatively weak seed $\sim 1-10$ mW.  At higher seed powers, in the range $10-100$ mW, we find that ASE can still compete for the gain with the signal, and significantly reduce amplifier efficiency. Once $S_s$ exceeds unity the effect of ASE becomes negligible. The effect of ASE on efficiency is studied in detail in \cref{broadband}, confirming the estimates given here.

Finally, when both pump and signal are saturated ($S_p,S_s \gg 1)$, most of the pump energy will be converted to amplification of the signal and the amplifier will operate near its maximum possible efficiency over the pump depletion length. Thus it is important to ensure that the seed is strong enough to saturate the signal for efficient operation, particularly in the multimode case where large core sizes result in lower intensity for a given power, and unsaturated speckle enables ASE and self-lasing. 

The effect of SE on amplifier efficiency is demonstrated in \cref{Efficiency} (a), beginning with a 10 $\upmu$m core diameter and 50 $\upmu$m cladding diameter fiber with 1 kW pump and 30 W signal, and sufficient length to absorb $95 \%$ of the pump. The core and cladding diameter is increased up to 200/1000 $\upmu$m, while keeping either the intensity constant or the total power constant. Dopant density $N_{tot}$ is scaled with 1/$r_{co}^2$ such that the total dopant number stays constant, and hence the pump absorption length remains similar across simulations. The number of excited modes is restricted to 20 for each core size supporting more - this is done both for consistency in simulation length and to avoid loss of efficiency from modes near cutoff for highly multimode fibers. The input signal is sufficient to suppress ASE in all cases, isolating SE as the efficiency loss mechanism.

Since scaling at constant intensity leaves the pump and signal saturation constant, as expected, the efficiency remains constant at almost $100\%$. However when the total power is kept constant, the signal intensity decreases below saturation and SE begins to degrade the efficiency significantly. The constant-power scaling reduces both the saturation of the signal and pump substantially, but up to the largest core size the pump remains saturated, so the main effect on efficiency comes from the decrease in the stimulated emission rate and hence increase in spontaneous emission rate relative to this.  



This decrease in amplifier efficiency with increasing core size due to SE is an effect which becomes increasingly relevant as the power-scaling potential of multimode fibers is pursued. Cladding-pumped fibers with large cladding sizes require very high pump power to strongly saturate the gain. If pump power is fixed then a large seed power can compensate with its own contribution to saturation, but for Yb this is highly inefficient and would require impractically large seed powers (see \cref{Threshold}).

\begin{figure*}[!t]
\includegraphics[width = \textwidth]{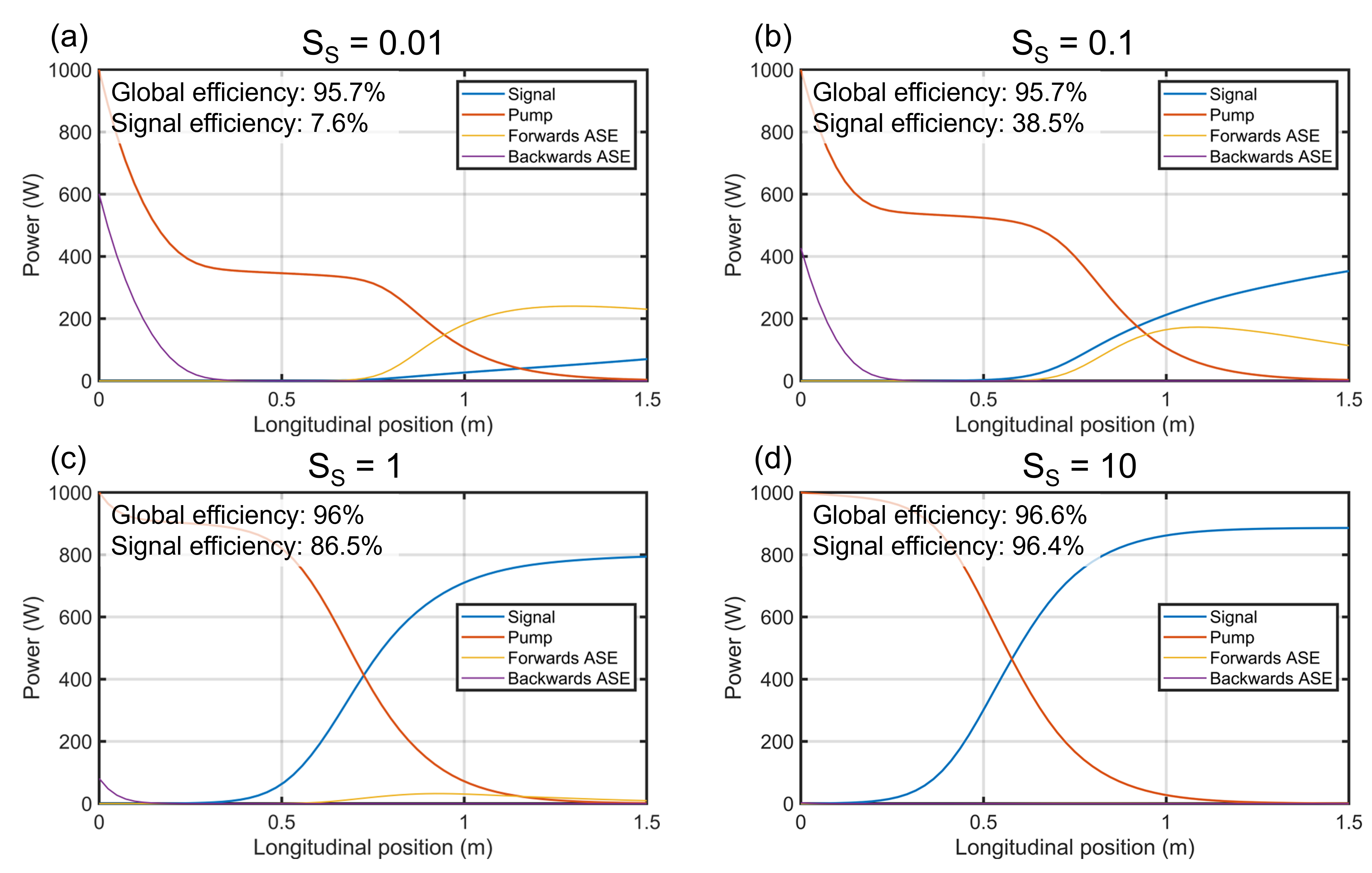}
\centering
\caption{Progression of pump power, signal power and backwards and forwards propagating ASE for an 80 $\upmu$m core diameter Yb-doped fiber with 1 kW pump input, and varying signal inputs. The input signal is expressed as the normalized signal power $S_S$ (as used in \cref{N?}), where for the given fiber parameters, $P_{sat,S}=36$ mW. This input signal power is divided equally amongst the first 20 modes of the 173 guided modes in the fiber, to avoid highly multimode effects such as high mode-dependent loss of modes near the cutoff.}
\label{ASESuppression}
\end{figure*}

\cref{Efficiency}(b) presents results similar to the red curve in (a), this time comparing a co- and counterpumped amplifier. These cases correspond to the input of the pump from z=0 (i.e. propagating with the signal) and z=L (i.e. propagating against the signal) respectively. The modeling of a counterpumped amplifier is performed using the shooting method \cite{osborne1969shooting}, the details of which are discussed in the Supplementary Information.

For this figure, other than the pumping configuration, we have used the same parameters as \cref{Efficiency}(a). In all cases the input signal is sufficiently saturating the gain and suppressing ASE, hence any decrease in efficiency is due to SE and insufficient pump saturation. The results demonstrate that the difference in efficiency between these configurations is negligible, with the counterpumped case producing an efficiency less than 1\% smaller on average than that of co-pumped. Hence, when the input signal is sufficiently saturating the gain, there is no advantage or disadvantage to co- or counterpumping a MMF amplifier from an efficiency point of view. A description of the modeling of forwards and backwards propagating ASE in the co- and counterpumped cases is presented in the Supplementary Information.

\section{Broadband MMF amplifier model} \label{BroadbandSection}

In the previous section, both pump and signal light were considered as single-frequency. However, for many fiber amplifier applications, such as the study of nonlinear pulse propagation \cite{chen2023,lindberg2016pulse} or wavelength-multiplexed telecommunications \cite{bergano1996wdm,brackett1990wdm}, the spectrally-dependent and broadband nature of the light propagating through the fiber becomes significant and even crucial. Furthermore, ASE is an inherently broadband process which can interfere with intended amplifier behavior, regardless of whether the signal light of interest is narrow-linewidth or not. In this section, a broadband MMF amplifier model is presented, which takes into account the spectrally-dependent nature of rare-earth dopants, again focusing specifically on Yb.

\subsection{Details of the broadband model} \label{broadband}

In the presence of broadband light, the response of the gain medium can be given as
\begin{gather}
N_2(\mathbf{r})=\frac{R_a(\mathbf{r})}{R_a(\mathbf{r})+R_e\ (\mathbf{r})+1/\tau} N_{tot}(\mathbf{r}), \label{N2spect}
\end{gather}
where
\begin{gather}
R_i(\mathbf{r})=\int\frac{1}{h\nu}\sigma_i\ (\nu)I(\mathbf{r},\nu)d\nu \label{Ri}
\end{gather}
are the spectrally integrated absorption and emission contributions to the overall population inversion. By expressing the light as two delta-functions in frequency at $\nu_P$ and $\nu_S$, \cref{N2spect} reduces to \cref{N2}.
For high-power, single-frequency amplifier operation, it remains reasonable to assume such single-frequency dependence for the pump and signal. However, ASE remains inherently broadband, and \cref{N2spect,Ri} are required to describe its effects on amplifier operation. Furthermore, one may wish to model the signal as a finitely narrow source, exploring the spectral evolution of its lineshape, making it necessary to give the signal spectral width.

For the below results, we assume that the pump is single frequency ($\nu_P$) and incoherent (no speckle behavior), while the signal and ASE exist over a discrete set of evenly-spaced wavelengths $\lambda_j=\frac{c}{\nu_j}$, each defining the center wavelength of a bin of width $\Delta\lambda$. This bin width is chosen so as to reasonably resolve the spectral cross-section curves shown in Fig. 3 of the Supplementary Information. In the continuous case, \cref{Ri} can now be expressed as
\begin{gather}
R_i(\mathbf{r})=\frac{\sigma_{i,P}}{h\nu_P}\frac{P_P(z,\nu_P)}{A_{cl}}+\int_\nu\frac{1}{h\nu}\sigma_i(\nu)I_{tot}(\mathbf{r},\nu)d\nu , \label{Rfull}
\end{gather}
where $I_{tot}(\mathbf{r},\nu)=I_S(\mathbf{r},\nu)+I_{ASE}(\mathbf{r},\nu)$. In both converting from frequency to wavelength and converting from the continuous to discrete case, the integral in \cref{Rfull} can be expressed as following:
\begin{gather}
\frac{1}{hc}\sum_j\sigma_i(\lambda_j)I_{tot}(\mathbf{r},\lambda_j)\lambda_j .
\end{gather}
ASE is assumed to be broadband; it is generated at every frequency which has a non-zero emission cross-section. It is also generated everywhere within the pumped section of the fiber and propagates in both the forwards and backwards directions. Modeling backwards ASE requires using the self-consistent relaxation method; a discussion of our method for modeling backwards ASE (together with backwards propagating pump) is given in the Supplementary Information.

At each z-step, spontaneous emission (SE) is generated due to the finite upper state lifetime of the active ions \cite{desurvire2002erbium,kogelnik1964ASE}, according to \cref{SE}:
\begin{gather}
\frac{dI_{SE}(z,\nu)}{dz}=2h\nu\delta\nu\sigma_e(\nu)\int  N_2(\mathbf{r})\hat{i}_{SE}(\mathbf{r},\nu)d\mathbf{r}_\perp . \label{SE}
\end{gather}
Here, $\hat{i}_{SE}$ describes the unit intensity profile which SE takes while propagating. In previous works \cite{trinel2017theoretical,chen2023}, it is assumed that SE is coupled into fiber modes in an incoherent manner, that is, ASE modes add as intensities, not fields. However, as the ASE band is wide (THz-10s of THz), its spectral correlation time is orders of magnitude smaller than the modal group delay spread. This means relative modal phases for one frequency are decorrelated from adjacent frequencies within the ASE band, and the spatial profile becomes smeared out like the incoherent pump. Hence we assume that ASE takes a uniform transverse spatial profile. The SE generated over a z-step is added to the existing accumulation of SE which has experienced amplification in the same way as the signal, and hence becomes ASE. Therefore, the growth of ASE throughout the fiber can be described by
\begin{gather}
\frac{dP_{ASE}(z,\nu)}{dz}=\frac{2h\nu\delta\nu\sigma_e(\nu)}{A_{co}}\int d\mathbf{r}_\perp N_2(\mathbf{r})\nonumber\\+\frac{P_{ASE}}{A_{co}}\int d\mathbf{r}_\perp(\sigma_e(\nu)N_2(\mathbf{r})-\sigma_a(\nu)N_1(\mathbf{r})) . \label{dPASEdz}
\end{gather}
Hence, equations \cref{N2spect,Rfull,dPASEdz}, together with \cref{ModalGain2} evaluated at each wavelength $\lambda_i$ and \cref{PumpAbsorption}, describe the propagation of broadband, multimode light through the gain medium of a fiber amplifier.

\subsection{Results of the broadband MMF amplifier model}

\begin{figure*}[!t]
\includegraphics[width = \textwidth]{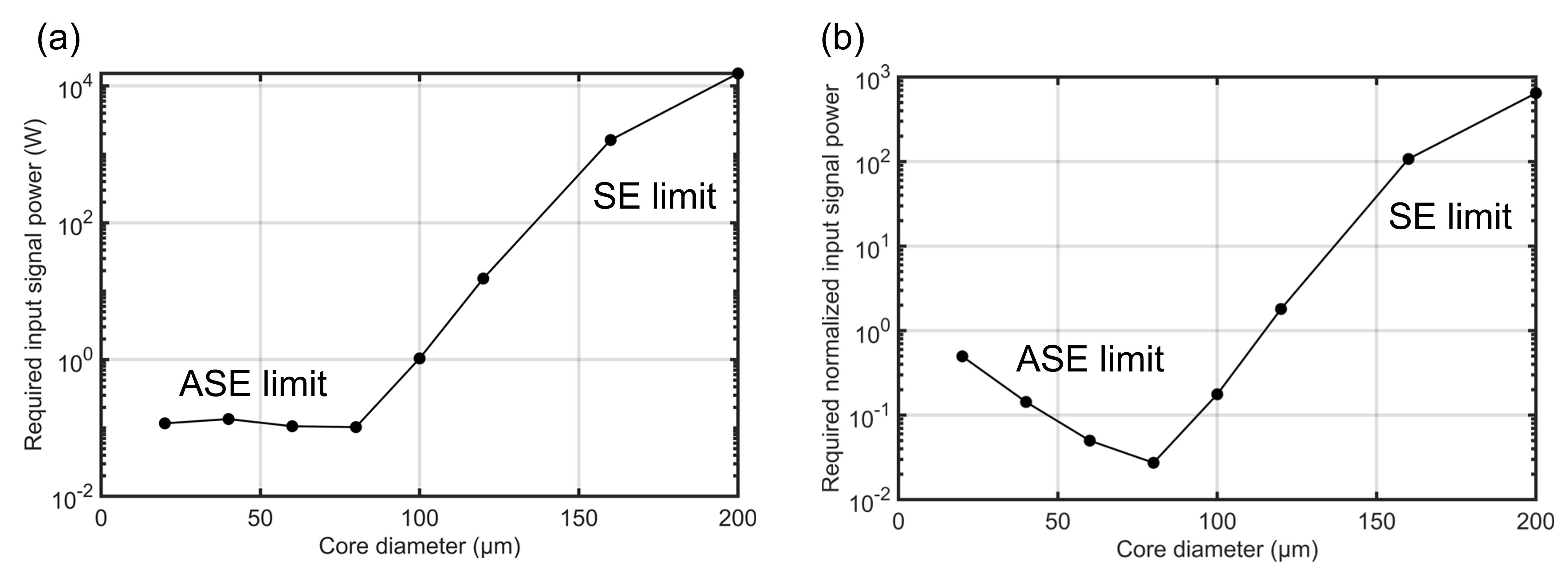}
\centering
\caption{(a) Input signal power and (b) normalized signal power ($S_S$) required to reach 95\% output signal efficiency at a length of 95\% pump absorption as a function of core diameter. The cladding diameter is set to 5$d_{co}$, hence for constant input pump power, the intensity and saturation decreases as $\sim d_{co}^{-2}$. The primary limits on amplifier efficiency for the regimes of small core, high pump saturation vs large core, low pump saturation are labelled as \textquote{ASE} limit and \textquote{SE} limit respectively.}
\label{Threshold}
\end{figure*}

In this section we evaluate the effect of ASE for Yb amplifiers using the broadband model. For Yb the main phenomenon is the suppression of ASE with increased input signal. For other rare-earth systems, such as Er, there are other spectral effects in the gain due to reabsorption effects, but we defer this analysis to future work. As alluded to in \cref{SingleFrequencySection}, sufficient signal saturation is required to suppress ASE. In general, the signal and ASE share the pump power not lost through isotropic SE emission, and one can define a `global' efficiency, which measures how efficiently pump power is converted into propagating light in the fiber. The global efficiency is mainly set by the level of pump saturation in relation to the SE rate, and the signal saturation determines how this efficiency is shared amongst signal band channels (signal + ASE).


As discussed in \cref{SingleFrequencySection}, the normalized signal power can be expressed as $S_S=P_S/P_{sat,S}$. For a saturated pump ($S_P>>1$) but weak signal, ASE is the main source of efficiency loss in a MMF amplifier. \cref{ASESuppression} gives amplifier behavior for $S_S=$ (a) 0.01, (b) 0.1, (c) 1 and (d) 10. Both the global and signal efficiencies are displayed. The signal efficiency is defined as before, while, as noted above, the global efficiency considers all channels which are drawing power from the pump (signal + forwards ASE + backwards ASE), with the Stokes efficiency for ASE considering its centroid wavelength of $\sim$1030 nm rather than 1064 nm for the signal.

As expected, increased signal power results in reduced ASE, with ASE becoming almost fully suppressed as $S_S$ approaches and exceeds unity. This is because the magnitude of SE generation depends on the upper-state lifetime $\tau$ and $S_P$, two parameters which here are constant. Hence the relative magnitude of SE generation depends only on the input signal. As such, the gain competition which ensues is also dependent only on signal input. The backwards propagating ASE serves to saturate the pump at the front of the fiber, and is suppressed at a similar rate with respect to input signal power as the forwards ASE.

In addition to demonstrating ASE suppression, \cref{ASESuppression} demonstrates that the global efficiency is set mainly by the level of pump saturation (given $f_a^p>>f_a^s$). While the absorbed power is shared between signal and ASE, we find that the sum of power in channels across the signal wavelength band will reach roughly the same efficiency regardless of how it is partitioned. Secondly, although signal absorption ($f_a^s$) for Yb is small, it is still non-negligible, especially for large input signal signal powers. Thus, despite each plot in \cref{ASESuppression} having the same input pump power, there is a small increase of 1\% in global efficiency due to the increase in input signal by three orders of magnitude from 0.01 to 10, however this is an incredibly inefficient method of increasing efficiency.

\cref{Threshold} gives the input signal required to reach 95\% signal efficiency as a function of core diameter in the case that cladding diameter is increased proportionally to core diameter, and hence pump saturation decreases with core diameter. Larger core sizes require stronger input pump power to suppress loss due to SE, while at smaller core sizes in which SE is already suppressed, input signal power must be large enough to suppress ASE. As seen in \cref{ASESuppression}, $S_S$ dropping below unity generally allows ASE to compete for gain with the signal and decrease signal efficiency.


For \cref{Threshold}, the cladding diameter is fixed such that $d_{cl}=5d_{co}$. Hence with a constant pump input of 1 kW, $S_P$ decreases as $\sim d_{co}^{-2}$. For core diameters up to $\sim$80 $\upmu$m, this pump saturation is still sufficient to set the global efficiency above the 95\% threshold by itself. In this regime, the signal power only needs to be strong enough to suppress ASE, hence being labeled the \textquote{ASE} limit and only requiring small input signal powers ($\sim$100 mW). For constant pump power, hence pump saturation decreasing as per $d_{co}^{-2}$, the signal saturation required to suppress ASE to the same level is also decreasing. This is because in the ASE limit, as the pump saturation decreases, so too does the population inversion. The result is a decreased rate of ASE and hence to suppress the ASE to similar levels, the required signal saturation decreases also.

However, for core diameters greater than $\sim$80 $\upmu$m, the pump saturation is not sufficient to set the global efficiency above the 95\% threshold on its own, hence high signal powers are required to make up the global efficiency lost to SE. Past the transition between the ASE and SE limits at approximately $d_{co}=80\:\upmu$m, the threshold input signal increases dramatically, due to the fact that $f_a^s>>f_a^p$ for Yb, and hence unreasonably high signal powers are needed to set the amplifier efficiency at 95\%.

These results highlight the need to consider pump and signal saturation levels when increasing fiber core size for reasons such as suppressing SBS or inducing higher order modes. While ASE suppression on its own only requires modest signal input powers and fixed saturation levels, SE itself becomes a harsh limit in the case that pump saturation is not sufficient. Large signal powers reaching a kW and above are required in such regimes, highlighting the need to dramatically increase pump power when entering the large core, multimode fiber amplifier regime.

\section{Discussion and conclusion}
We have presented a derivation and analysis of a field-based MMF amplifier model, applicable to highly-multimode, high-power fiber lasers, an application of such a model as yet unexplored in the literature. The derivation in \cref{ModalGainDerivation} allows insight and intuition into how the multimode character of the signal and noise in such amplifiers affects their properties and specifically their efficiency. In such systems the signal is typically narrowband, whereas the ASE noise is broadband, hence we have introduced both a `single-frequency' system of equations for modeling the signal and SE, and a broadband model with evolution equations appropriate to modeling ASE. Our formulation of this broadband model allows a computationally efficient treatment specific to the highly multimode regime, allowing us to analyze effects such as efficiency loss to ASE as a function of seed power, as well as the spectral shape of this ASE. The model is then applied to Yb-doped fiber amplifiers of varying core and cladding dimensions.

The single-frequency results depict modal gain and amplifier efficiency in the context of high-power Yb fiber amplifier operation. Such fiber amplifiers are increasingly relevant for applications such as the generation/suppression of non-linear effects and wavefront shaping/beamshaping techniques. We find that such an amplifier can operate at a large fraction of its Stokes efficiency up to $\sim 80\:\upmu$m core diameter with a 1 kW pump and 30 W seed. The pump will need to be stronger to have high efficiency beyond that core diameter. We show that a non-saturated pump can be somewhat compensated by a larger seed, but this is an inefficient and impractical approach for Yb. By employing our broadband model we show that power loss to ASE for Yb is a negligible effect for seeds above $\sim$100s of mW. However this may not be the case for other dopants such as Er, where reabsorption of the signal is much stronger than in Yb. Our model suggests Er will have significant differences from Yb due to the differing spectroscopic characteristics between the two ions. Such effects are planned to be studied in future work.
\\
\section*{Acknowledgements}
This work was funded by the US Air Force Office of Scientific Research grant FA-9550-20-1-0160. Stephen Warren-Smith is supported by an Australian Research Council (ARC) Future Fellowship (FT200100154). This work is supported by the Australian Research Council (ARC) under CE230100016. Darcy Smith acknowledges the support of the American Australian Association for the support in his research and thanks the Yale Quantum Institute for the use of their facilities. The authors acknowledge and thank Linh V. Nguyen of the Future Industries Institute, Adelaide University, for his stimulating discussions and useful contributions to this work.


\bibliography{main}

@article{trinel2017theoretical,
  title={Theoretical study of gain-induced mode coupling and mode beating in few-mode optical fiber amplifiers},
  author={Trinel, Jean-Baptiste and Le Cocq, Guillaume and Quiquempois, Yves and Andresen, Esben Ravn and Vanvincq, Olivier and Bigot, Laurent},
  journal={Optics express},
  volume={25},
  number={3},
  pages={2377--2390},
  year={2017},
  publisher={Optica Publishing Group}
}

@article{chen2023,
author = {Yi-Hao Chen and Henry Haig and Yuhang Wu and Zachary Ziegler and Frank Wise},
journal = {J. Opt. Soc. Am. B},
number = {10},
pages = {2633--2642},
publisher = {Optica Publishing Group},
title = {Accurate modeling of ultrafast nonlinear pulse propagation in multimode gain fiber},
volume = {40},
month = {Oct},
year = {2023},
url = {https://opg.optica.org/josab/abstract.cfm?URI=josab-40-10-2633},
doi = {10.1364/JOSAB.500586},
}

@article{agrawal1990amplification,
  title={Amplification of ultrashort solitons in erbium-doped fiber amplifiers},
  author={Agrawal, Govind P},
  journal={IEEE Photonics technology letters},
  volume={2},
  number={12},
  pages={875--877},
  year={1990},
  publisher={IEEE}
}

@ARTICLE{giles1991EDFAs,
  author={Giles, C.R. and Desurvire, E.},
  journal={Journal of Lightwave Technology}, 
  title={Modeling erbium-doped fiber amplifiers}, 
  year={1991},
  volume={9},
  number={2},
  pages={271-283},
  doi={10.1109/50.65886}}

@article{paschotta1997ytterbium,
  title={Ytterbium-doped fiber amplifiers},
  author={Paschotta, Rudiger and Nilsson, Johan and Tropper, Anne C and Hanna, David C},
  journal={IEEE Journal of quantum electronics},
  volume={33},
  number={7},
  pages={1049--1056},
  year={1997},
  publisher={IEEE}
}

@article{kan2012intensity,
author = {Qiongyue Kang and Ee-Leong Lim and Yongmin Jung and Jayanta K. Sahu and Francesco Poletti and Catherine Baskiotis and Shaif-ul Alam and David J. Richardson},
journal = {Opt. Express},
number = {19},
pages = {20835--20843},
publisher = {Optica Publishing Group},
title = {Accurate modal gain control in a multimode erbium doped fiber amplifier incorporating ring doping and a simple LP01 pump configuration},
volume = {20},
month = {Sep},
year = {2012},
url = {https://opg.optica.org/oe/abstract.cfm?URI=oe-20-19-20835},
doi = {10.1364/OE.20.020835},
}

@article{gomes2022focussing,
author = {Andr\'{e} D. Gomes and Sergey Turtaev and Yang Du and Tom\'{a}\v{s} \v{C}i\v{z}m\'{a}r},
journal = {Opt. Express},
number = {7},
pages = {10645--10663},
publisher = {Optica Publishing Group},
title = {Near perfect focusing through multimode fibres},
volume = {30},
month = {Mar},
year = {2022},
url = {https://opg.optica.org/oe/abstract.cfm?URI=oe-30-7-10645},
doi = {10.1364/OE.452145},
}

@article{florentin2017activeMMF,
   author = {Florentin, Raphael and Kermene, Vincent and Benoist, Joel and Desfarges-Berthelemot, Agnès and Pagnoux, Dominique and Barthélémy, Alain and Huignard, Jean-Pierre},
   title = {Shaping the light amplified in a multimode fiber},
   journal = {Light: Science \& Applications},
   volume = {6},
   number = {2},
   pages = {e16208-e16208},
   ISSN = {2047-7538},
   DOI = {10.1038/lsa.2016.208},
   url = {https://doi.org/10.1038/lsa.2016.208},
   year = {2017},
   type = {Journal Article}
}

@article{rothe2024beamshaping,
author = {Rothe, Stefan and Wisal, Kabish and Chen, Chun-Wei and Ercan, Mert and Jesacher, Alexander and Stone, A.Douglas and Cao, Hui},
year = {2024},
month = {12},
pages = {131405},
title = {Output beam shaping of a multimode fiber amplifier},
volume = {577},
journal = {Optics Communications},
doi = {10.1016/j.optcom.2024.131405}
}

@article{jauregui2013HPFLs,
   author = {Jauregui, Cesar and Limpert, Jens and Tünnermann, Andreas},
   title = {High-power fibre lasers},
   journal = {Nature Photonics},
   volume = {7},
   number = {11},
   pages = {861-867},
   ISSN = {1749-4893},
   DOI = {10.1038/nphoton.2013.273},
   url = {https://doi.org/10.1038/nphoton.2013.273},
   year = {2013},
   type = {Journal Article}
}

@article{dawson2008HPFLs,
   author = {Dawson, Jay W. and Messerly, Michael J. and Beach, Raymond J. and Shverdin, Miroslav Y. and Stappaerts, Eddy A. and Sridharan, Arun K. and Pax, Paul H. and Heebner, John E. and Siders, Craig W. and Barty, C. P. J.},
   title = {Analysis of the scalability of diffraction-limited fiber lasers and amplifiers to high average power},
   journal = {Optics Express},
   volume = {16},
   number = {17},
   pages = {13240-13266},
   DOI = {10.1364/OE.16.013240},
   url = {https://opg.optica.org/oe/abstract.cfm?URI=oe-16-17-13240},
   year = {2008},
   type = {Journal Article}
}

@article{zervas2014HPFLs,
   author = {Zervas, M. N. and Codemard, C. A.},
   title = {High Power Fiber Lasers: A Review},
   journal = {IEEE Journal of Selected Topics in Quantum Electronics},
   volume = {20},
   number = {5},
   pages = {219-241},
   ISSN = {1558-4542},
   DOI = {10.1109/JSTQE.2014.2321279},
   year = {2014},
   type = {Journal Article}
}

@article{hecht18HPFLs,
author = {Jeff Hecht},
journal = {Opt. Photon. News},
number = {10},
pages = {30--37},
publisher = {Optica Publishing Group},
title = {High-Power Fiber Lasers},
volume = {29},
month = {Oct},
year = {2018},
url = {https://www.optica-opn.org/abstract.cfm?URI=opn-29-10-30},
doi = {10.1364/OPN.29.10.000030},
}

@article{wellmann19gravwave,
author = {Felix Wellmann and Michael Steinke and Fabian Meylahn and Nina Bode and Benno Willke and Ludger Overmeyer and J\"{o}rg Neumann and Dietmar Kracht},
journal = {Opt. Express},
number = {20},
pages = {28523--28533},
publisher = {Optica Publishing Group},
title = {High power, single-frequency, monolithic fiber amplifier for the next generation of gravitational wave detectors},
volume = {27},
month = {Sep},
year = {2019},
url = {https://opg.optica.org/oe/abstract.cfm?URI=oe-27-20-28523},
doi = {10.1364/OE.27.028523},
}

@article{carter19welding,
author = {Richard Carter},
journal = {American Ceramic Society Bulletin},
number = {4},
pages = {20--27},
publisher = {The American Ceramic Society},
title = {From concept to industry: Ultrafast laser welding},
volume = {100},
month = {May},
year = {2021},
url = {https://ceramics.org/wp-content/uploads/2021/04/May-2021_Feature.pdf},
}

@article{jauregui2020tmi,
   author = {Jauregui, Cesar and Stihler, Christoph and Limpert, Jens},
   title = {Transverse mode instability},
   journal = {Advances in Optics and Photonics},
   volume = {12},
   number = {2},
   pages = {429-484},
   DOI = {10.1364/AOP.385184},
   url = {https://opg.optica.org/aop/abstract.cfm?URI=aop-12-2-429},
   year = {2020},
   type = {Journal Article}
}

@article{zervas2018tmi,
    author = {Zervas, Michalis N.},
    title = {Transverse-modal-instability gain in high power fiber amplifiers: Effect of the perturbation relative phase},
    journal = {APL Photonics},
    volume = {4},
    number = {2},
    pages = {022802},
    year = {2018},
    month = {12},
    issn = {2378-0967},
    doi = {10.1063/1.5050523},
    url = {https://doi.org/10.1063/1.5050523},
    eprint = {https://pubs.aip.org/aip/app/article-pdf/doi/10.1063/1.5050523/14568654/022802\_1\_online.pdf},
}

@article{smith2011tmi,
author = {Arlee V. Smith and Jesse J. Smith},
journal = {Opt. Express},
number = {11},
pages = {10180--10192},
publisher = {Optica Publishing Group},
title = {Mode instability in high power fiber amplifiers},
volume = {19},
month = {May},
year = {2011},
url = {https://opg.optica.org/oe/abstract.cfm?URI=oe-19-11-10180},
doi = {10.1364/OE.19.010180},
}

@article{kobyakov2010SBS,
author = {Andrey Kobyakov and Michael Sauer and Dipak Chowdhury},
journal = {Adv. Opt. Photon.},
number = {1},
pages = {1--59},
publisher = {Optica Publishing Group},
title = {Stimulated Brillouin scattering in optical fibers},
volume = {2},
month = {Mar},
year = {2010},
url = {https://opg.optica.org/aop/abstract.cfm?URI=aop-2-1-1},
doi = {10.1364/AOP.2.000001},
}

@article{wolff2021SBS,
author = {C. Wolff and M. J. A. Smith and B. Stiller and C. G. Poulton},
journal = {J. Opt. Soc. Am. B},
number = {4},
pages = {1243--1269},
publisher = {Optica Publishing Group},
title = {Brillouin scattering---theory and experiment: tutorial},
volume = {38},
month = {Apr},
year = {2021},
url = {https://opg.optica.org/josab/abstract.cfm?URI=josab-38-4-1243},
doi = {10.1364/JOSAB.416747},
}

@article{agrawal1989MI,
  title = {Modulation instability induced by cross-phase modulation in optical fibers},
  author = {Agrawal, Govind P. and Baldeck, P. L. and Alfano, R. R.},
  journal = {Phys. Rev. A},
  volume = {39},
  issue = {7},
  pages = {3406--3413},
  numpages = {0},
  year = {1989},
  month = {Apr},
  publisher = {American Physical Society},
  doi = {10.1103/PhysRevA.39.3406},
  url = {https://link.aps.org/doi/10.1103/PhysRevA.39.3406}
}

@article{dupiol2017MI,
author = {R. Dupiol and A. Bendahmane and K. Krupa and J. Fatome and A. Tonello and M. Fabert and V. Couderc and S. Wabnitz and G. Millot},
journal = {Opt. Lett.},
number = {17},
pages = {3419--3422},
publisher = {Optica Publishing Group},
title = {Intermodal modulational instability in graded-index multimode optical fibers},
volume = {42},
month = {Sep},
year = {2017},
url = {https://opg.optica.org/ol/abstract.cfm?URI=ol-42-17-3419},
doi = {10.1364/OL.42.003419},
}

@article{wisal2024nonlinear,
author = {Wisal, Kabish and Chen, Chun-Wei and Kuang, Zeyu and Miller, Owen and Cao, Hui and Stone, A.Douglas},
year = {2024},
month = {12},
pages = {1663-1672},
title = {Optimal input excitations for suppressing nonlinear instabilities in multimode fibers},
volume = {11},
journal = {Optica},
doi = {10.1364/OPTICA.533712}
}

@article{chen2023wavefrontshaping,
author = {Chen, Chun-Wei and Nguyen, Linh and Wisal, Kabish and Wei, Shuen and Warren-Smith, Stephen and Henderson-Sapir, Ori and Schartner, Erik and Ahmadi, Peyman and Ebendorff-Heidepriem, Heike and Stone, A.Douglas and Ottaway, David and Cao, Hui},
year = {2023},
month = {11},
pages = {},
title = {Mitigating stimulated Brillouin scattering in multimode fibers with focused output via wavefront shaping},
volume = {14},
journal = {Nature Communications},
doi = {10.1038/s41467-023-42806-1}
}

@article{wisal2024TMI,
author = {Wisal, Kabish and Chen, Chun-Wei and Cao, Hui and Stone, A.Douglas},
year = {2024},
month = {06},
pages = {},
title = {Theory of transverse mode instability in fiber amplifiers with multimode excitations},
volume = {9},
journal = {APL Photonics},
doi = {10.1063/5.0206859}
}

@article{wisal2024sbs,
author = {Wisal, Kabish and Warren-Smith, Stephen and Chen, Chun-Wei and Cao, Hui and Stone, A.Douglas},
year = {2024},
month = {09},
pages = {},
title = {Theory of Stimulated Brillouin Scattering in Fibers for Highly Multimode Excitations},
volume = {14},
journal = {Physical Review X},
doi = {10.1103/PhysRevX.14.031053}
}

@article{lindberg2016pulse,
   author = {Lindberg, Robert and Zeil, Peter and Malmström, Mikael and Laurell, Fredrik and Pasiskevicius, Valdas},
   title = {Accurate modeling of high-repetition rate ultrashort pulse amplification in optical fibers},
   journal = {Scientific Reports},
   volume = {6},
   number = {1},
   pages = {34742},
   ISSN = {2045-2322},
   DOI = {10.1038/srep34742},
   url = {https://doi.org/10.1038/srep34742},
   year = {2016},
   type = {Journal Article}
}

@ARTICLE{brackett1990wdm,
  author={Brackett, C.A.},
  journal={IEEE Journal on Selected Areas in Communications}, 
  title={Dense wavelength division multiplexing networks: principles and applications}, 
  year={1990},
  volume={8},
  number={6},
  pages={948-964},
  doi={10.1109/49.57798}}

@ARTICLE{bergano1996wdm,
  author={Bergano, N.S. and Davidson, C.R.},
  journal={Journal of Lightwave Technology}, 
  title={Wavelength division multiplexing in long-haul transmission systems}, 
  year={1996},
  volume={14},
  number={6},
  pages={1299-1308},
  doi={10.1109/50.511662}}

@ARTICLE{kogelnik1964ASE,
  author={Kogelnik, H. and Yariv, A.},
  journal={Proceedings of the IEEE}, 
  title={Considerations of noise and schemes for its reduction in laser amplifiers}, 
  year={1964},
  volume={52},
  number={2},
  pages={165-172},
  doi={10.1109/PROC.1964.2805}}

@book{snyder1983optical,
  title={Optical Waveguide Theory},
  author={Snyder, A.W. and Love, J.},
  isbn={9780412099502},
  lccn={lc83007463},
  year={1983},
  publisher={Springer US}
}

@book{agrawal2019nonlinear,
   author = {Agrawal, Govind P.},
   title = {Nonlinear Fiber Optics},
   publisher = {Academic Press},
   edition = {6th},
   year = {2019},
   type = {Book}
}

@book{siegman1986lasers,
   author = {Siegman, A.E.},
   title = {Lasers},
   publisher = {University Science Books},
   ISBN = {9780935702118},
   year = {1986},
   type = {Book}
}

@book{hecht2017optics,
  title={Optics},
  author={Hecht, E.},
  isbn={9780133977226},
  lccn={2016000110},
  url={https://books.google.com/books?id=ZarLoQEACAAJ},
  year={2017},
  publisher={Pearson Education, Incorporated}
}

@book{desurvire2002erbium,
   author = {Desurvire, Emmanuel},
   title = {Erbium-Doped Fiber Amplifiers},
   publisher = {Wiley},
   address = {New Jersey, USA},
   year = {2002},
   type = {Book}
}

@article{liu2016lidar,
	author = {{Liu, Jiqiao} and {Zhu, Xiaopeng} and {Diao, Weifeng} and {Zhang, Xin} and {Liu, Yuan} and {Bi, Decang} and {Jiang, Liyuan} and {Shi, Wei} and {Zhu, Xiaolei} and {Chen, Weibiao}},
	title = {All-Fiber Airborne Coherent Doppler Lidar to Measure Wind Profiles},
	DOI= "10.1051/epjconf/201611910002",
	url= "https://doi.org/10.1051/epjconf/201611910002",
	journal = {EPJ Web of Conferences},
	year = 2016,
	volume = 119,
	pages = "10002",
}

@ARTICLE{kaushal2017defense,
  author={Kaushal, Hemani and Kaddoum, Georges},
  journal={IEEE Access}, 
  title={Applications of Lasers for Tactical Military Operations}, 
  year={2017},
  volume={5},
  number={},
  pages={20736-20753},
  doi={10.1109/ACCESS.2017.2755678}}

@article{krupa2017selfcleaning,
   author = {Krupa, K. and Tonello, A. and Shalaby, B. M. and Fabert, M. and Barthélémy, A. and Millot, G. and Wabnitz, S. and Couderc, V.},
   title = {Spatial beam self-cleaning in multimode fibres},
   journal = {Nature Photonics},
   volume = {11},
   number = {4},
   pages = {237-241},
   ISSN = {1749-4893},
   DOI = {10.1038/nphoton.2017.32},
   url = {https://doi.org/10.1038/nphoton.2017.32},
   year = {2017},
   type = {Journal Article}
}

@ARTICLE{ke2014sbs,
  author={Ke, Wei-Wei and Wang, Xiao-Jun and Tang, Xuan},
  journal={IEEE Journal of Selected Topics in Quantum Electronics}, 
  title={Stimulated Brillouin Scattering Model in Multi-Mode Fiber Lasers}, 
  year={2014},
  volume={20},
  number={5},
  pages={305-314},
  doi={10.1109/JSTQE.2014.2303256}}

@article{osborne1969shooting,
title = {On shooting methods for boundary value problems},
journal = {Journal of Mathematical Analysis and Applications},
volume = {27},
number = {2},
pages = {417-433},
year = {1969},
issn = {0022-247X},
doi = {https://doi.org/10.1016/0022-247X(69)90059-6},
url = {https://www.sciencedirect.com/science/article/pii/0022247X69900596},
author = {M.R Osborne}
}

@article{chen2023TMI,
author = {Chen, Chun-Wei and Wisal, Kabish and Eliezer, Yaniv and Stone, A.Douglas and Cao, Hui},
year = {2023},
month = {05},
pages = {e2217735120},
title = {Suppressing transverse mode instability through multimode excitation in a fiber amplifier},
volume = {120},
journal = {Proceedings of the National Academy of Sciences of the United States of America},
doi = {10.1073/pnas.2217735120}
}

@article{chen2025,
author = {Chen, Chun-Wei and Wisal, Kabish and Fink, Mathias and Stone, A.Douglas and Cao, Hui},
year = {2025},
month = {04},
pages = {839-845},
title = {Output control of dissipative nonlinear multimode amplifiers using spacetime symmetry mapping},
volume = {21},
journal = {Nature Physics},
doi = {10.1038/s41567-025-02853-5}
}

@ARTICLE{wright2018computational,
  author={Wright, Logan G. and Ziegler, Zachary M. and Lushnikov, Pavel M. and Zhu, Zimu and Eftekhar, M. Amin and Christodoulides, Demetrios N. and Wise, Frank W.},
  journal={IEEE Journal of Selected Topics in Quantum Electronics}, 
  title={Multimode Nonlinear Fiber Optics: Massively Parallel Numerical Solver, Tutorial, and Outlook}, 
  year={2018},
  volume={24},
  number={3},
  pages={1-16},
  doi={10.1109/JSTQE.2017.2779749}}

@article{rothe2025wavefrontshaping,
author = {Rothe, Stefan and Chen, Chun-Wei and Ahmadi, Peyman and Lee, KyeoReh and Wisal, Kabish and Ercan, Mert and Vigne, Nathan and Stone, A.Douglas and Cao, Hui},
year = {2025},
month = {10},
pages = {173-177},
title = {Wavefront shaping enables high-power multimode fiber amplifier with output focus},
volume = {390},
journal = {Science (New York, N.Y.)},
doi = {10.1126/science.ady2226}
}
\end{document}